%% file: full_draft.tex
\documentclass[sigconf]{acmart}

\usepackage[utf8]{inputenc} 
\usepackage[T1]{fontenc}    
\usepackage{hyperref}       
\usepackage{url}            
\usepackage{booktabs}       
\usepackage{amsfonts}       
\usepackage{nicefrac}       
\usepackage{microtype}      
\usepackage{xcolor}         
\usepackage[T1]{fontenc}
\usepackage{graphicx}
\usepackage[normalem]{ulem}

\setcopyright{acmlicensed}\acmConference[SIGIR '24]{Proceedings of the 47th International ACM SIGIR Conference on Research and Development in Information Retrieval}{July 14--18, 2024}{Washington, DC, USA} \acmBooktitle{Proceedings of the 47th International ACM SIGIR Conference on Research and Development in Information Retrieval (SIGIR '24), July 14--18, 2024, Washington, DC, USA}
\acmDOI{10.1145/3626772.3657677} \acmISBN{979-8-4007-0431-4/24/07}

\begin{document}

\title{JPEC: A Novel Graph Neural Network for Competitor Retrieval in Financial Knowledge Graphs}
\author{Wanying Ding}
\email{wanying.ding@jpmchase.com}
\affiliation{%
  \institution{JPMorgan Chase \& Co.}
  \city{Palo Alto}
  \state{California}
  \country{USA}
}

\author{Manoj Cherukumalli}
\email{manoj.cherukumalli@jpmchase.com}
\affiliation{%
  \institution{JPMorgan Chase \& Co.}
  \city{Palo Alto}
  \state{California}
  \country{USA}
}

\author{Santosh Chikoti}
\email{santosh.chikoti@jpmchase.com}
\affiliation{%
  \institution{JPMorgan Chase \& Co.}
  \city{Jersey City}
  \state{New Jersey}
  \country{USA}
}

\author{Vinay K. Chaudhri}
\email{vinay.chaudhri@jpmchase.com}
\affiliation{%
  \institution{JPMorgan Chase \& Co.}
  \city{Palo Alto}
  \state{California}
  \country{USA}
}

\begin{abstract}
Knowledge graphs have gained popularity for their ability to organize and analyze complex data effectively. When combined with graph embedding techniques, such as graph neural networks (GNNs), knowledge graphs become a potent tool in providing valuable insights. This study explores the application of graph embedding in identifying competitors from a financial knowledge graph. Existing state-of-the-art(SOTA) models face challenges due to the unique attributes of our knowledge graph, including directed and undirected relationships, attributed nodes, and minimal annotated competitor connections. To address these challenges, we propose a novel graph embedding model, JPEC(\underline{J}PMorgan \underline{P}roximity \underline{E}mbedding for \underline{C}ompetitor Detection), which utilizes graph neural network to learn from both first-order and second-order node proximity together with vital features for competitor retrieval. JPEC had outperformed most existing models in extensive experiments, showcasing its effectiveness in competitor retrieval.
\end{abstract}



\maketitle
\section{Introduction}
\input{introduction}

\section{Problem and Proposed Method}
\label{sc:problem and solution}
\input{problem_definition}
\input{methods}

\section{Experimental Settings}
\label{sc:experiment}
\input{dataset}
\input{baselines}

\section{Results and Discussions}
\input{result}

\section{Related Work}
\input{related_work}

\section{Conclusion}
In summary, this paper introduces JPEC, a novel graph embedding model designed for competitor detection from a financial knowledge graph. Leveraging two orders of node proximity and essential features, JPEC surpasses most SOTA graph embedding models and human queries from various extensive experiments. Our evaluations of various graph embedding models offer valuable insights into their strengths and limitations for real-world competitor detection tasks. JPEC represents a significant contribution to the field, showcasing the potential of knowledge graphs and graph embedding techniques in unveiling patterns within complex networks in practical business scenarios.

\bibliographystyle{abbrv}
\bibliography{reference}  
\end{document}

%% file: introduction.tex
Competitor retrieval is one of the most crucial use cases for financial organizations. Traditionally, it is mostly driven by multiple manual tasks involving collecting data and converting factors like revenue, products, pricing, marketing, and industry distributions. While manually gathered market data offer vital insights, they have limitations in terms of applicability and scalability. On the other hand, knowledge graphs can provide competitive clues by revealing meaningful connections, such supply-chain, between companies. Combined with graph embedding techniques, knowledge graphs can offer a structured and efficient approach for automatic and intelligent competitor retrieval. However, most SOTA graph embedding methods are sub-optimal for our task due to the complex structure of a real world knowledge graph(described in Section\ref{sc:problem and solution}).  This paper introduces a novel graph neural network, JPEC, for competitor detection from a financial knowledge graph with various types of edges but limited labeled data.

%% file: problem_definition.tex
\subsection{Problem Definition}
In our knowledge graph $\mathcal{G}=(\mathcal{V},\mathcal{S},\mathcal{C},\mathcal{X})$, each node in the node-set $\mathcal{V}$ represents a real-world company, and $\mathcal{X}$ contains attributes associated with each node. The directed edge set $\mathcal{S}$ signifies supply chain connections between companies, while the undirected edge set $\mathcal{C}$ denotes mutual competitor relationships. Notably, our knowledge graph lacks numerous competitor edges, resulting in a significantly smaller volume for $\mathcal{C}$ compared to $\mathcal{S}$. Our objective is to leverage the limited competitor edges, combined with the extensive company node attributes and supply chain graph structure, to identify additional competitors for a given company.

%% file: methods.tex
\begin{figure*}[t]
\centering 
\includegraphics[width=0.6\textwidth]{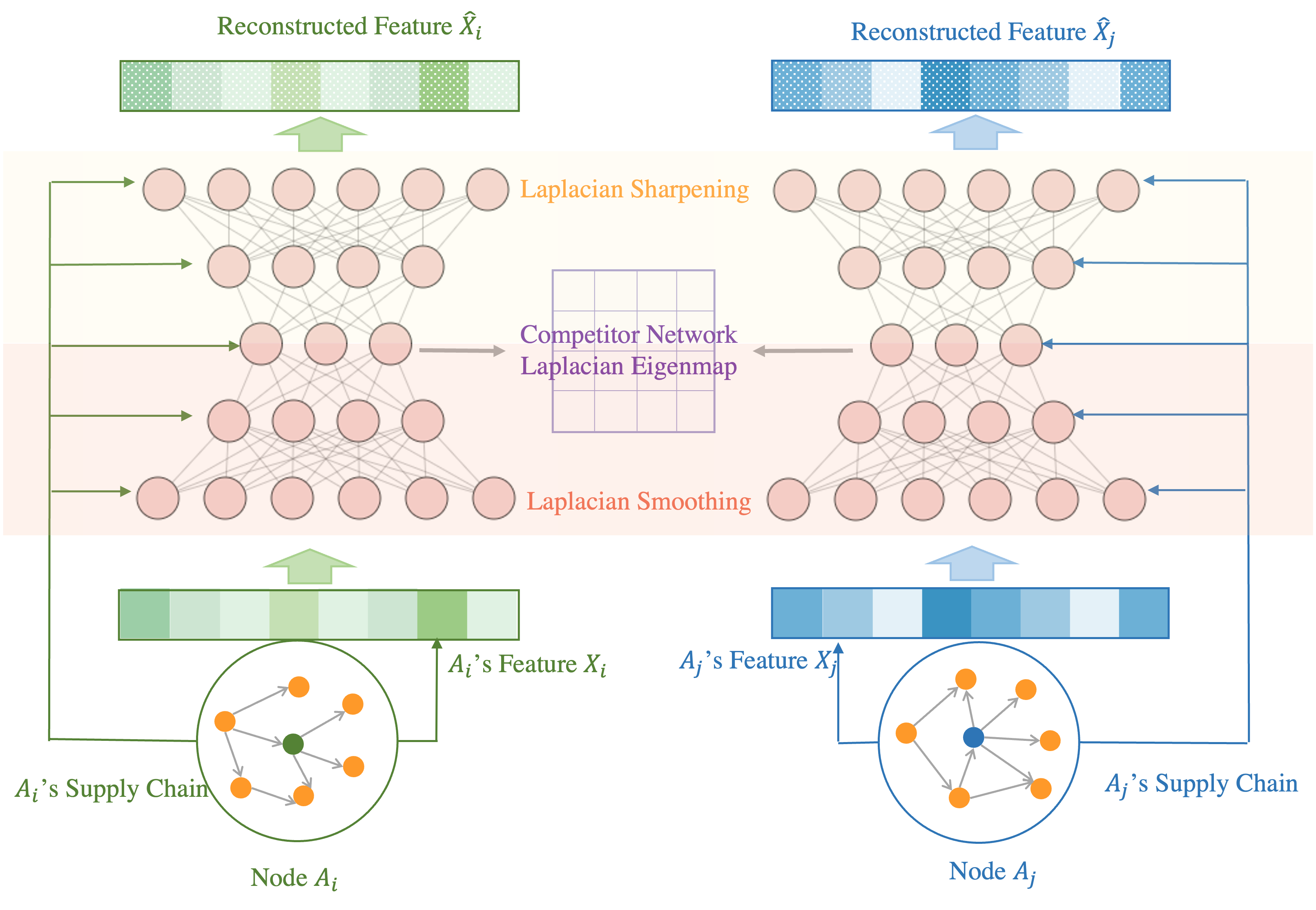}
 \caption{The Model Framework for JPEC}
\label{fg:model}
\end{figure*}
\subsection{Model Description}
We introduce JPEC (shown as in  Figure\ref{fg:model}) model for competitor detection. This model leverages two orders of proximity for effective competitor pattern capture.

\subsubsection{First Order Proximity: Laplacian Eigenmap on Competitor Network}
Although competitor edges are sparse in our graph, learning from these connections is natural and essential. We employ first-order proximity to characterize the local connection and use it as supervised information that constrains the similarity of latent representations between pairs of competitors\cite{wang2016structural}. The weight $w_{ij}$ along the competitor edge between node $i$ and $j$ is $1$ if they are known competitors, $-1$ if they are non-competitors, otherwise $0$.

We applied Laplacian Eigenmap\cite{belkin2003laplacian} to enforce the first proximity. Laplacian Eigenmap aims at constructing a Laplacian matrix $L$ to ensure nodes connected as competitors stay as close as possible after embedding. We designed two Laplacian Eigenmaps to learn from positive samples (Equation \ref{eq:pos_loss}) and negative samples (Equation \ref{eq:neg_loss}) respectively. It is worth noting that Laplacian Eigenmap only accepts non-negative weights, so we used the opposite edge weight for non-competitors, $-w^{-}_{i,j}$, in Equation \ref{eq:neg_loss}. 

\begin{equation}
    \mathcal{L}_{pos} = \sum_{i,j=1}^{N} w^{+}_{i,j} ||y_{i}-y_{j}|| = 2tr(Y^{T}L^{+}Y)
    \label{eq:pos_loss}
\end{equation}
\begin{equation}
    \mathcal{L}_{neg} = \sum_{i,j=1}^{N} -w^{-}_{i,j} ||y_{i}-y_{j}|| = 2tr(Y^{T}L^{-}Y)
    \label{eq:neg_loss}
\end{equation}
where $L^{+}$ is the Laplacian matrix for nodes in positive samples, and $L^{-}$ is for nodes in negative samples, $tr(\cdot)$ is the trace of a matrix, and $Y$, indicating the node embeddings. We'll elaborate on how to generate $Y$ in the next section. 

Finally, we utilized a pairwise ranking loss function (Equation \ref{eq:first_loss}) to minimize the distance between positive pairs and simultaneously maximizing the distance between negative pairs, where $m$ denotes the margin, which is a hyper-parameter controlling the desired separation space between positive and negative pairs. 
\begin{equation}
    \mathcal{L}_{1st} = \mathcal{L}_{pos}+max(0,m-\mathcal{L}_{neg})
    \label{eq:first_loss}
\end{equation}

\subsubsection{Second Order Proximity: Directed GCN Autoencoder on Supply Chain Network}
In this section, we explain how we obtain node embeddings $Y$ for the aforementioned objective function. Since each node has associated attributes, GCN is a straightforward option to utilize and learn graph structure and attributes simultaneously. GCN is naturally designed for undirected graphs, and we change the GCN's propagation function $\tilde{D}^{-1/2}\tilde{A}\tilde{D}^{-1/2}$ to $\tilde{D}^{-1}\tilde{A}$ , to apply it into a directed supply-chain graph\cite{schlichtkrull2018modeling,shi2019skeleton}. By changing the normalization function, the propagation rule of GCN can be rewritten as Equation \ref{eq:laplacian_smoothing}
\begin{equation}
Y^{(l+1)} = \sigma(\tilde{D}^{-1}\tilde{A}Y^{(l)}W^{(l)})
\label{eq:laplacian_smoothing}
\end{equation}
where $\tilde{A}=A+I$ is the adjacency matrix. $I$ is the identity matrix, $\tilde{D}$ is the degree matrix,  and $W^{(l)}$ is layer-specific trainable weight matrix, $\sigma(\cdot)$ denotes an activation function. $Y^{(l)} \in \mathbb{R}^{N \times D}$ is hidden representations in the $l^{th}$ layer. 

Since many competitor edges are missing in our graph, a decoder is necessary to enhance the model's ability to extract information from the supply chain graph. Since GCN is a Laplacian smoothing process, we employ a Laplacian sharpening process\cite{park2019symmetric} to reverse the encoding process. Similarity, we made adaptions to accommodate the characteristics to a directed graph(shown as Equation \ref{eq:laplacian_sharpening}). 
\begin{equation}
    Y^{(m+1)} = \sigma ((\hat{D}^{-1}\hat{A})Y^{(m)}W^{(m)})
    \label{eq:laplacian_sharpening}
\end{equation}
The loss function for the second order proximity is to minimize the difference between the original node feature vectors and the reconstructed ones, which can be formulated as Equation \ref{eq:second_loss}:
\begin{equation}
    \mathcal{L}_{2nd} = ||X-\hat{X}||_{2}^{2}
    \label{eq:second_loss}
\end{equation}
The ultimate objective function of our model integrates the loss function derived from both the first-order and second-order proximity, and can be mathematically represented as Equation \ref{eq:loss}.
\begin{equation}
\mathcal{L} = \mathcal{L}_{1st}+\beta \mathcal{L}_{2nd}+\lambda \mathbf{W}^{2}
\label{eq:loss}
\end{equation}
where $\beta$ is a hyper parameter to balance the first-order and second-order losses, $\lambda \mathbf{W}^{2}$ is the regularization term. 
\begin{figure*}[h]
\centering 
\includegraphics[width=0.7\textwidth]{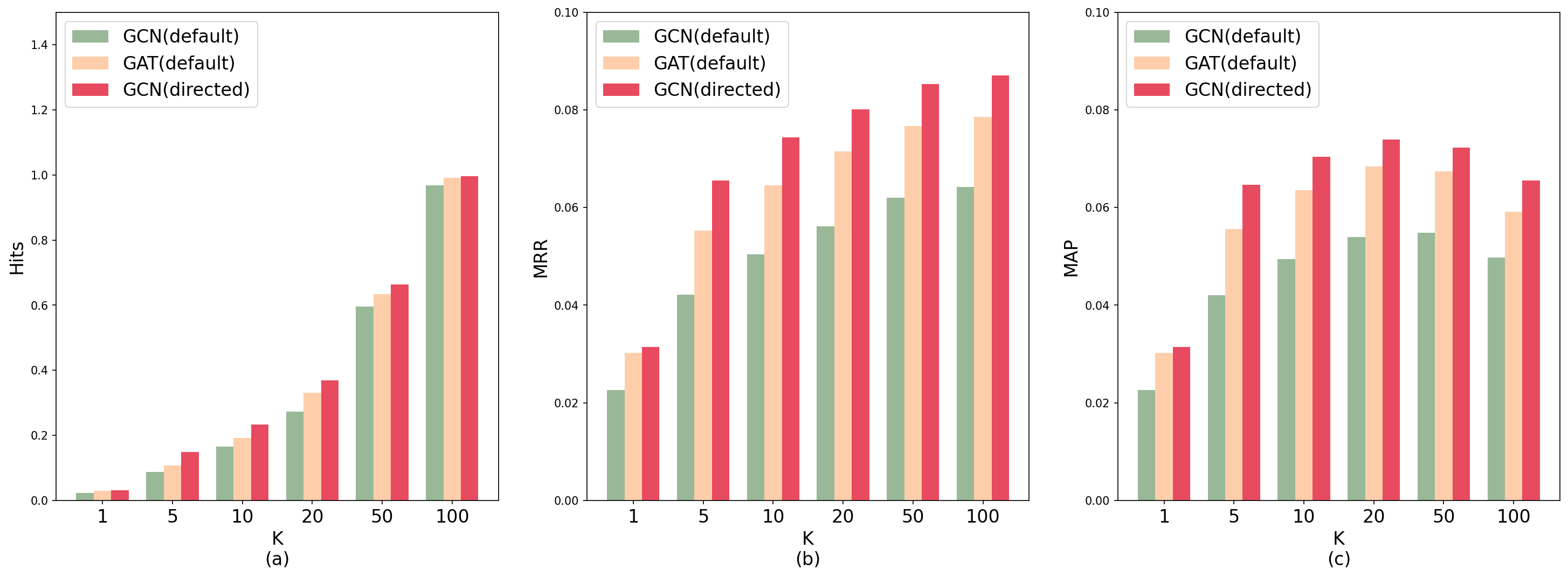}
 \caption{Comparison among GCN, GAT, and DGCN}
\label{fg:gcn_compare}
\end{figure*}

%% file: dataset.tex
\subsection{Datasets}
We possess a large-scaled financial knowledge graph offering a comprehensive overview of financial entities (e.g., companies, investors, bankers) and their relationships (e.g., supply chain, investment). From this knowledge graph, we extracted a subgraph $G$ consisting of $133,812$ company nodes and $611,812$ associated supply chain edges. Within this sample graph, $39,755$ nodes, equivalent to 30\% of the total nodes, have competitors, resulting in a total of $213,071$ competitor edges. To facilitate a comprehensive experiment, we generated two evaluation datasets: a regular test dataset $R$ and a zero-shot test dataset $Z$.

\subsubsection{Zero-Shot Test Dataset Preparation} 
With 70\% of companies' competitors absent in our knowledge graph, assessing a model's generalization capability for predicting competitors not in the graph becomes crucial. We selected a subset of $7,952$ nodes(20\% of all nodes) and extracted COMPETE\_WITH edges around them. We then removed all COMPETE\_WITH connections between these nodes and the rest of the graph to ensure these nodes are unseen in the training competitor data. But their supply chain relationships are retained and utilized for competitor prediction. In testing, we exclusively consider companies with a minimum of 5 competitors, yielding a subset of $201$ companies.

\subsubsection{Regular Test Dataset Preparation}
We performed random sampling on the remaining dataset after creating the zero-shot test dataset. In contrast to the zero-shot test dataset, the regular test dataset retains all nodes but randomly removes some COMPETE\_WITH edges from the graph. It worth noting that a node is still visible in the training competitor data since it still has other competitors in addition to the removed ones.  We extracted 20\% of the edges out as test data. Similarly, we curated a subset of companies with a minimum of 5 competitors, resulting in a final set of $795$ companies for the regular test.

We randomly designated certain unconnected nodes in the competitor network as non-competitors. This method may introduce errors since some sampled nodes may be competitors with missing COMPETE\_WITH edge. Nevertheless, we are dedicated to refining our negative sampling methodology to mitigate such issues.


%% file: baselines.tex
\subsection{Tasks and Evaluation Metrics}
In the context of competitor retrieval, our objective is to identify the competitors of a given company. We compare models' performances with three ranking metrics. \textbf{Hits} calculates the rate of correct items appearing in each instance list's top $K$ entries. \textbf{MRR} stands for Mean Reciprocal Rank. It tries to measure where the first correct item is in the predicted list. \textbf{MAP} stands for Mean Average Precision, which measures the average precision across a set of items. 

\subsection{Baselines and Settings}
We employ two types of baselines: Human and Graph Embedding. For the Human baseline, we collaborated with our business team to manually detect competitors from the knowledge graph, translating their expertise into proprietary graph queries. This approach, constrained to utilizing information solely within the knowledge graph, yielded satisfactory outcomes as confirmed through a review of sample results. Regarding graph embedding, we evaluated various methods, including structure embedding (Node2Vec, Metapath2Vec, TransE, and SDNE), attributed embeddings (GCN, GAT, DGCN, and DGAE), and hetergeneous methods (HAN and KBGAT).

%% file: result.tex
\subsection{Default GCN vs. Directed GCN}
Before examining the model performance on two tasks, we briefly validate the directed GCN used in this paper. We implemented three methods: GCN, GAT, and DGCN. Figure \ref{fg:gcn_compare} shows DGCN's superior performance compared to the others. GCN's limitations stem from its reliance on symmetric adjacency matrices, while GAT performs better than GCN but worse than DGCN due to it neglects of edge directions, potentially causing errors.

\begin{figure*}[h]
\centering 
\includegraphics[width=0.75\textwidth]{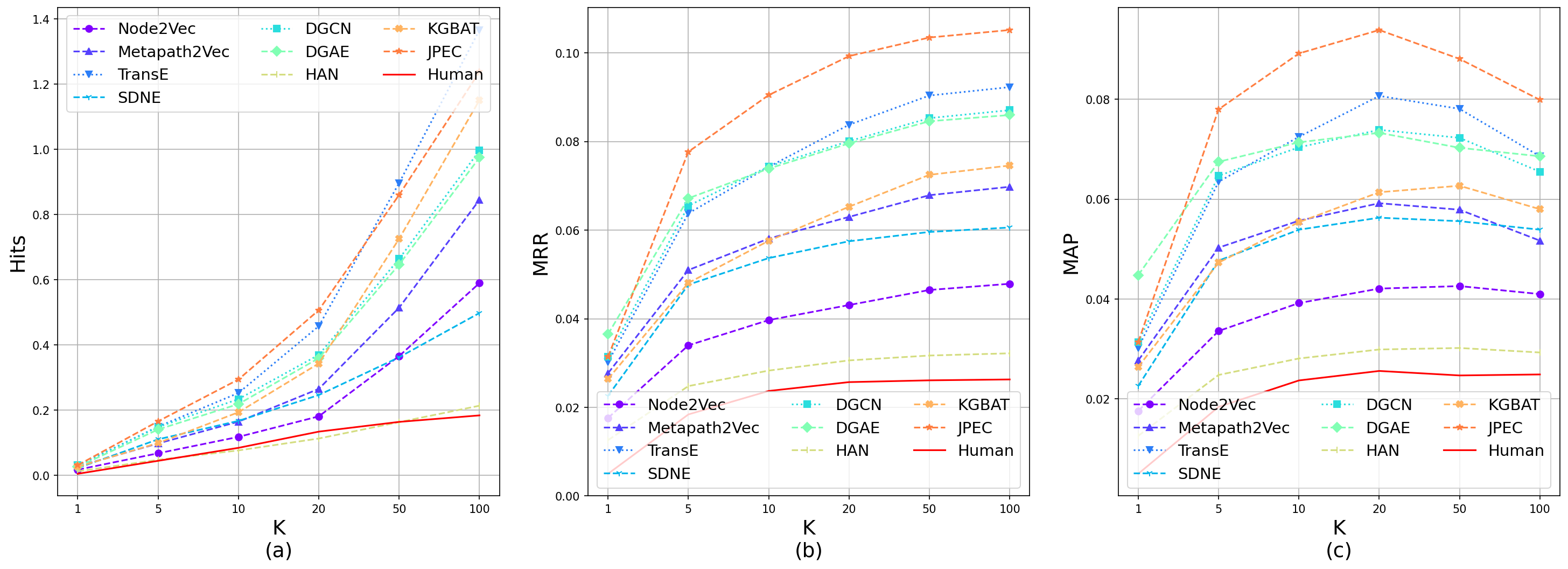}
 \caption{Models' Performances on Regular Test}
\label{fg:regular_rank}
\end{figure*}

\begin{figure*}[h]
\centering 
\includegraphics[width=0.75\textwidth]{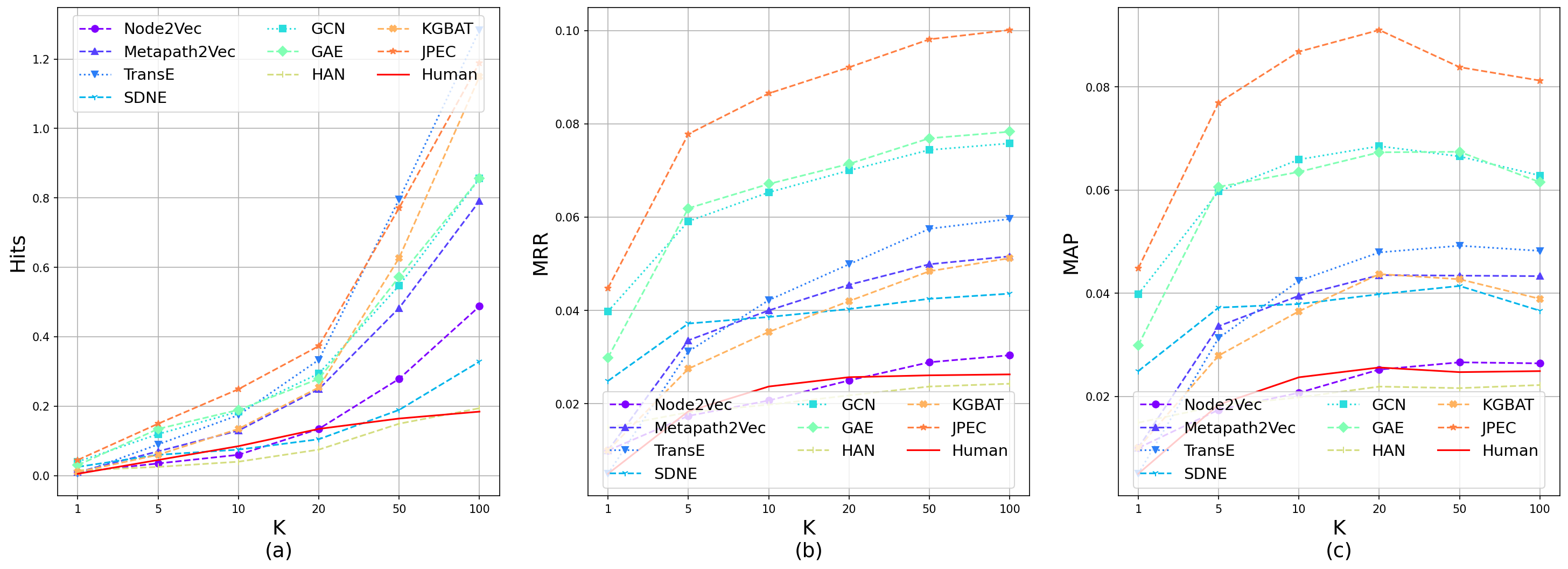}
\caption{Models' Performances on Zero-Shot Test}
\label{fg:zero_rank}
\end{figure*}

\subsection{Discussion on Competitor Retrieval}
Competitor Retrieval is a very important task in our business operations. Given a company, our goal is to retrieve a roaster of potential competitors. Figure \ref{fg:regular_rank} displays the performance of each model evaluated on our regular test datasets, whereas Figure \ref{fg:zero_rank} exhibits the performance on our zero-shot test dataset. The red solid line in each figure represents the performance of human queries. The to-be-ranked candidate pool is all the $133, 812$ companies. 

\subsubsection{Regular Testing}
Figure \ref{fg:regular_rank} shows most machine learning-based methods outperforming human queries, highlighting graph topology's importance in competitor detection. HAN's limitations stem from significant information loss due to its extraction method, while DGCN, DGAE, and KBGAT excel by considering node attributes. KBGAT's shortcomings arise from noises accumulated from n-hop information aggregation, whereas DGAE's performance is limited by its unlearnable decoder. JPEC and TransE rank high in identifying competitors, but TransE struggles to prioritize true competitors over non-competitors in terms of much lower MRR and MAP.

\subsubsection{Zero-Shot Testing}
Figure \ref{fg:zero_rank} tests each model's generalization capability through zero-shot testing. The red line, representing human queries, remains consistent across both figures. Structure embedding methods like Node2Vec, MetaPath2Vec, SDNE, and TransE perform worse than attributed embedding methods like DGCN and DGAE. TransE's significant decline indicates a cold-start problem, difficult to make accurate predictions on unseen nodes. Attributed embedding methods, considering both graph structure and node attributes, exhibit more stability, similar to recommender systems.

 JPEC consistently outperforms other methods on both regular and zero-shot test datasets, demonstrating its effectiveness and superiority in competitor discovery.

%% file: related_work.tex
Graph embedding techniques, including Graph Neural Networks (GNN), transform nodes, edges, and attributes into lower-dimensional vector spaces while preserving graph properties. This paper applies graph embedding to competitor detection, categorizing methods into Structure Embedding, Attributed Embedding, and Heterogeneous Embedding. Structure Embedding models, such as DeepWalk\cite{perozzi2014deepwalk}, Node2Vec\cite{grover2016node2vec}, Metapath2Vec\cite{dong2017metapath2vec}, TransE\cite{wang2014knowledge}, and SDNE\cite{wang2016structural}, focus on graph topology. Attributed Embedding, including GCN\cite{zhang2019graph}, GAE\cite{kipf2016variational}, and GALA\cite{park2019symmetric}, emphasizes node attributes. Heterogeneous Embedding, with methods like HAN \cite{wang2019heterogeneous} and KBGAT\cite{nathani2019learning}, represents diverse node and edge types. The proposed JPEC model, inspired by SDNE\cite{wang2016structural} and GALA\cite{park2019symmetric}, employs a directed graph convolutional network (DGCN) and Laplacian sharpening for semi-supervised competitor detection, outperforming existing models in the financial domain.